\documentclass{PoS}

\title{The prompt-afterglow connection in Gamma-Ray Bursts: a comprehensive statistical analysis of Swift X-ray light-curves}

\ShortTitle{The GRB X-ray afterglow}

\author{\speaker{Raffaella Margutti}\\
        Harvard-Smithsonian Center for Astrophysics, 60 Garden Street, Cambridge, MA 02138, USA\\
        E-mail: \email{rmargutti@cfa.harvard.edu}}

\author{E. Zaninoni\\
        INAF - Osservatorio Astronomico di Brera, via Bianchi 46, I-23807 Merate (LC), Italy\\
        University of Padova, Physics \& Astronomy Dept. Galileo Galilei, via Marzolo 8, I-35131 Padova, Italy\\}

\author{M. G. Bernardini\\
        INAF - Osservatorio Astronomico di Brera, via Bianchi 46, I-23807 Merate (LC), Italy\\}

\author{G. Chincarini\\
        INAF - Osservatorio Astronomico di Brera, via Bianchi 46, I-23807 Merate (LC), Italy\\
        University of Milano Bicocca, Physics Dept., p.zza della Scienza 3, I-20126 Milano, Italy\\} 
        
\author{on behalf of the Swift-XRT team\\}

\abstract{We present a comprehensive statistical analysis of \emph{Swift} X-ray 
light-curves of Gamma-Ray Bursts (GRBs), with more than 650 GRBs. Two
questions drive this effort: (1) Does the X-ray emission retain any kind of memory of the 
prompt phase? (2) Where is the dividing line between long and short GRBs?
We show that short GRBs decay faster, are less luminous and less energetic than 
long GRBs, but are interestingly characterized by very similar intrinsic absorption.
Our analysis reveal the existence of a number of relations that link the X-ray to prompt 
parameters in long GRBs; short GRBs are outliers of the majority of these 2-parameter
relations. Here we concentrate on a 3-parameter ($E_{\rm{pk}}-E_{\rm{\gamma,iso}}-E_{\rm{X,iso}}$) 
scaling that is shared by the GRB class as a whole (short GRBs, long GRBs and X-ray Flashes -XRFs):
interpreted in terms of emission efficiency, this scaling may imply that GRBs with high $E_{\rm{pk}}$ 
are more efficient during their prompt emission.}

\FullConference{Gamma-Ray Bursts 2012 Conference -GRB2012,\\
		May 07-11, 2012\\
		Munich, Germany}

\begin{document}

\section{Introduction}
The extremely fast  re-pointing capabilities of the \emph{Swift} 
spacecraft  enabled the GRB community to sample the X-ray emission that follows
the prompt phase in exquisite detail starting as early as $\sim 60$ s after the trigger.
After $\sim 7$ years of operation (and more than 700 GRBs observed by the X-ray telescope
on board \emph{Swift}  -XRT), we have now the possibility to discuss  the properties 
of the X-ray light-curves of GRBs from  a statistical perspective.  We analyze more than 650 
GRBs (38 are short GRBs) detected by \emph{Swift} -XRT during the first 6 years of operation. 
For the subsample of 437 GRBs with complete light-curve (i.e. promptly re-pointed by
\emph{Swift}-XRT and for which we have been able to follow the fading of the source
down to a factor 5-10 from the background limit)
we are able to constrain the properties of their X-ray light-curves 
(decaying slopes, temporal break times, fluxes, fluences) together with the properties
(duration, peak flux, statistical significance, fluence) of the X-ray flares superimposed.
Our approach benefits from the largest sample of GRBs with redshift measurement analyzed
in the literature (165 GRBs) in the common rest-frame energy band 0.3-30 keV:
this gives us the possibility to measure their \emph{intrinsic}
properties (time-scales, energetics). Furthermore, we combine the results from our
analysis in the X-rays with the prompt $\gamma$-ray emission parameters from the literature
\cite{Sakamoto11, Amati08, Nava08}, and look for correlations that bridge the gap between
the prompt and afterglow emission. 

The results from our X-ray flare project can be found in \cite{Margutti10, Margutti11, Margutti11b,
Chincarini10, Bernardini11}. 
Here we focus on the major results we 
obtained from the study of the smoothly decaying X-ray afterglow component. We refer
the reader to \cite{Margutti12} for a detailed description of our findings.

\section{Major Results}
\subsection{Short vs. long GRB X-ray afterglow properties}

\begin{figure}
\vskip -0.0 true cm
\centering
\includegraphics[scale=0.5]{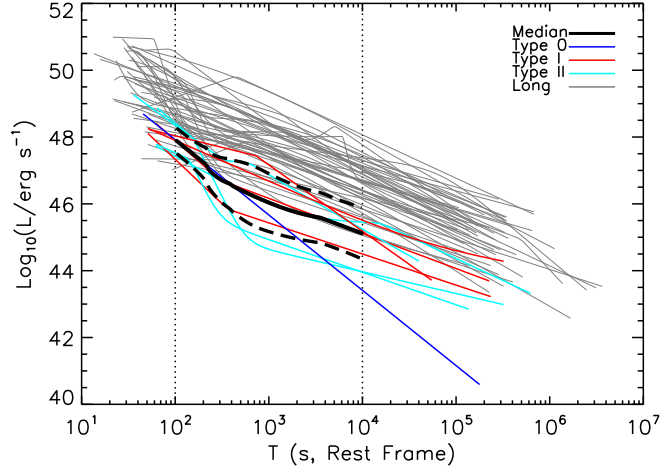}
\caption{Grey lines: 0.3-30 keV (rest-frame) best fitting profiles of 77 long GRBs of our sample
with complete light-curve (i.e. promptly re-pointed by \emph{Swift} and whose follow-up
has not been truncated). Color lines: sample of 9 short GRBs satisfying the same criteria. Different
colors refer to different light-curve morphological types as defined by \cite{Margutti12}. The black thick
line marks the median light-curve for the short sample. }  
\label{Fig:shortaft}
\end{figure}

\begin{figure}
\vskip -0.0 true cm
\centering
\includegraphics[scale=0.3]{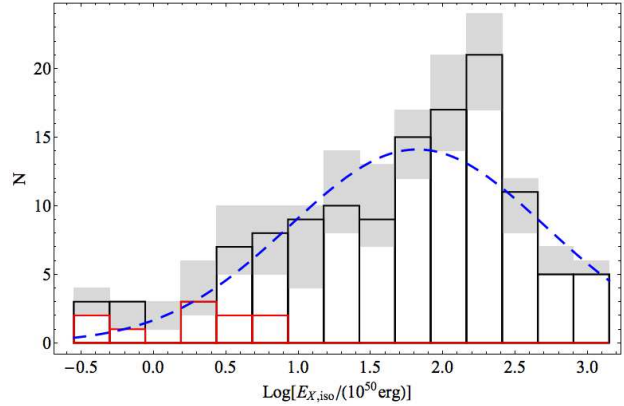}
\caption{0.3-30 keV (rest-frame) isotropic energy emitted after the
prompt emission is over. Black (red) solid line: long (short) GRBs. The grey
area is derived using Monte Carlo simulations and marks the 99\% confidence interval.
Blue dashed line: best-fitting Gaussian profile with central value $E_{\rm{X,iso}}\sim 6\times 10^{51}\,\rm{erg}$.  }  
\label{Fig:Etotx}
\end{figure}

\begin{figure}
\vskip -0.0 true cm
\centering
\includegraphics[scale=0.4]{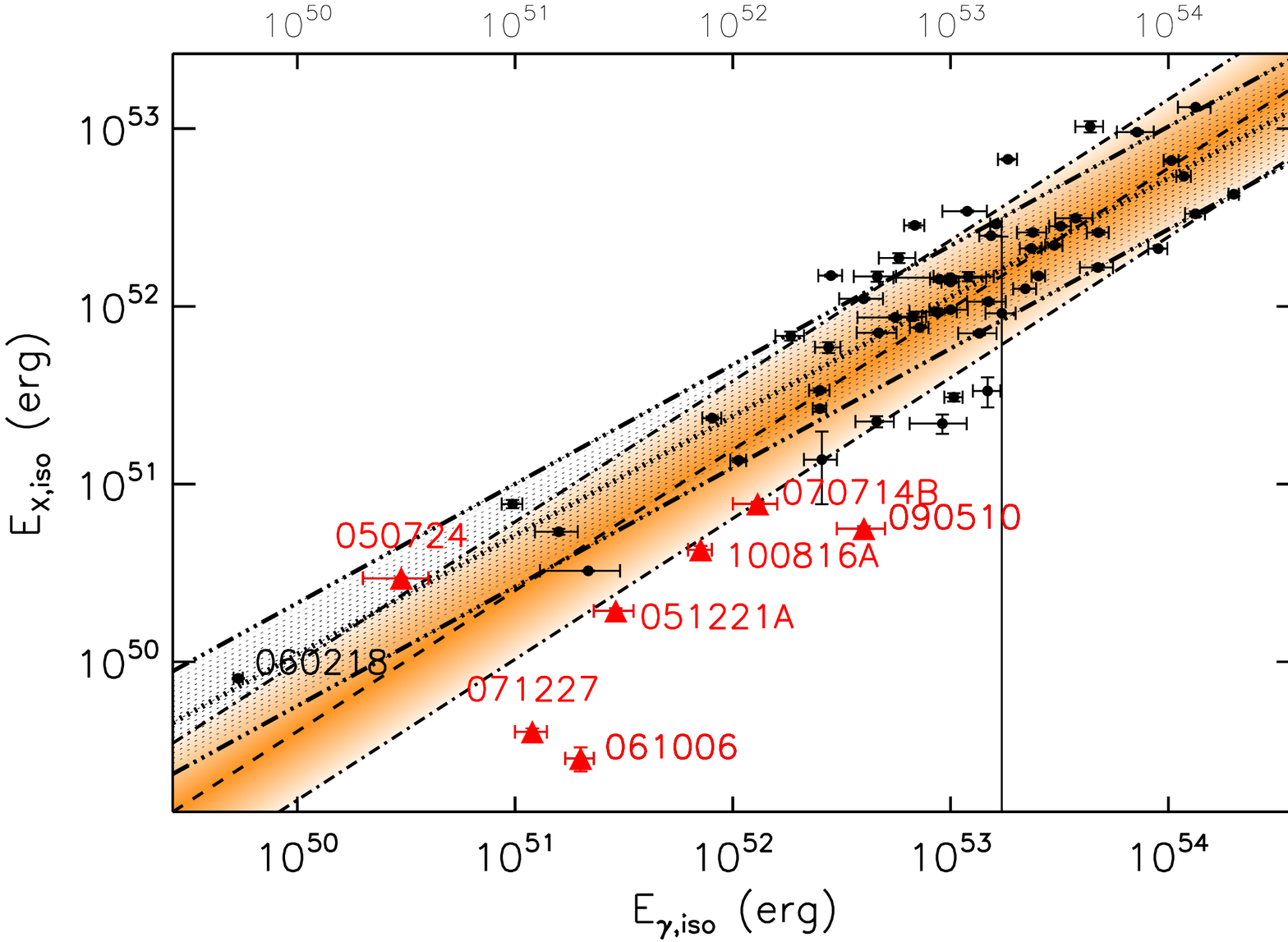}
\caption{Isotropic X-ray (0.3-30 keV, rest frame) energy emitted after the prompt emission is over 
vs. prompt $\gamma$-ray ($1-10^4$) energy. Black (red) points: long (short) GRBs. Dashed line:
best-fitting relation for the entire sample: $E_{\rm{X,iso}}\propto E_{\rm{\gamma,iso}}^{0.8}$. 
Excluding the short GRBs, we obtain $E_{\rm{X,iso}}\propto E_{\rm{\gamma,iso}}^{0.7}$. The shaded
areas mark the 68\% confidence region around the best-fitting relations.}  
\label{Fig:EtotxEiso}
\end{figure}

\begin{figure}
\vskip -0.0 true cm
\centering
\includegraphics[scale=0.7]{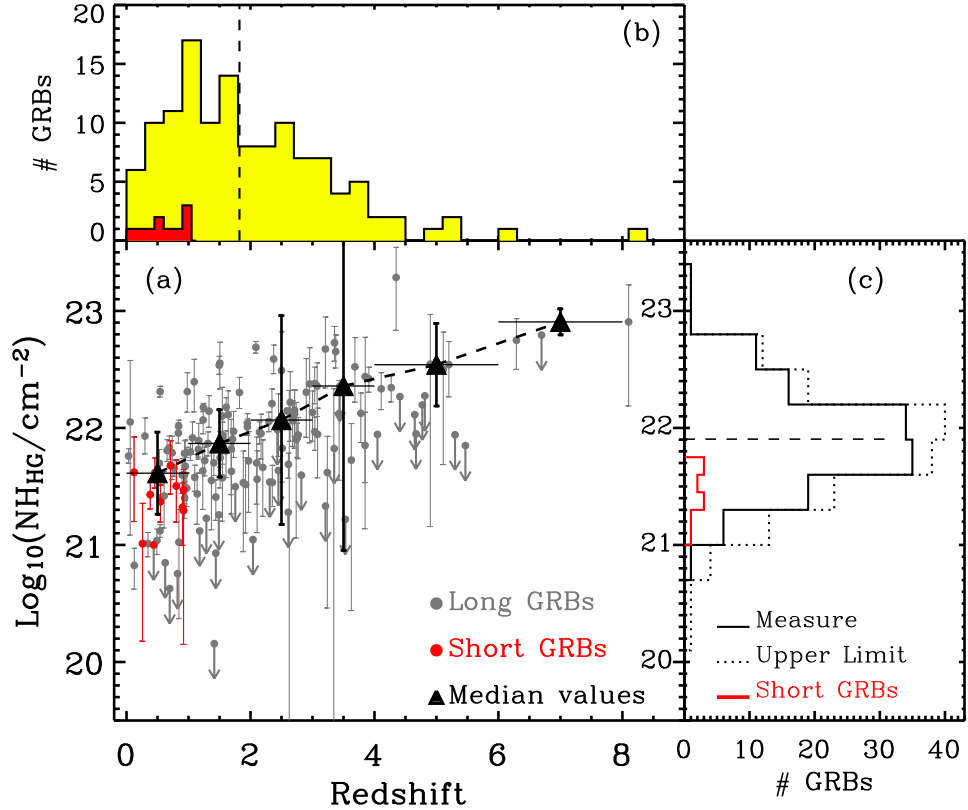}
\caption{(a): Intrinsic neutral hydrogen absorption vs. redshift for long and short GRBs (grey and red dots, respectively). 90\% upper limits are marked with arrows. Median $NH_{\rm{HG}}$ values in different redshift bins are indicated with filled triangles.  In panels (b) and (c) a dashed line indicates the median value for the entire distributions ($z=1.82$, $NH_{\rm{HG}} = 10^{21.8}\rm{cm^{-2}}$).}  
\label{Fig:NH}
\end{figure}

Short GRBs are traditionally classified  from their properties during the 
prompt $\gamma$-ray phase: a short duration ($T_{90}<2$ s), a hard $\gamma$-ray 
emission  and a negligible
spectral time lag during the prompt emission are considered suggestive of a ``short"-GRB nature.
Do short GRBs show a distinct behavior during their X-ray afterglow as well? Our study 
reveals that:\\
(1) Short GRB X-ray afterglows are less luminous of a factor $\sim10-30$ when compared
to long GRB afterglows (Fig. \ref{Fig:shortaft}). However, Figure \ref{Fig:shortaft} demonstrates that 
the two samples slightly overlap, especially at early ($t_{\rm{rest}}<300$ s) times, when short GRBs
typically show a luminosity comparable to the  low luminosity edge of the long GRB distribution.\\
(2) On average, short GRB afterglows decay faster (average decay in the 
rest-frame time interval $10^2-10^4$ s $\propto t^{-1}$
vs. $\propto t^{-1.3}$ for long and short GRBs, respectively. The uncertainty associated to the
decay index is $\lesssim0.05$).\\
(3) The average 0.3-30 keV (rest-frame) isotropic energy released during the afterglow 
of a short GRB is $\sim100$ times lower than a long GRB (see Fig. \ref{Fig:Etotx}):
in particular, $E_{\rm{X,iso}}^{short}< 10^{51}\,\rm{erg}$. \\
(4) For long GRBs, the X-ray energy emitted after the prompt emission correlates
with the isotropic equivalent prompt $\gamma$-ray energy (Fig. \ref{Fig:EtotxEiso}). With the exception of
GRB\,050724, short GRBs fall off the $E_{\rm{X,iso}}$ vs. $E_{\rm{\gamma,iso}}$ relation established
by long GRBs: when compared to long GRBs, short GRBs emit less energy in the X-rays than expected. \\
(5) Short GRBs map the low end of the intrinsic neutral hydrogen $NH_{\rm{HG}}$ distribution of GRBs
(Fig. \ref{Fig:NH}, panel $c$), with an average absorption $NH_{\rm{HG}}^{short} = 10^{21.4}\rm{cm^{-2}}$
(the median value for the entire distribution is $NH_{\rm{HG}} = 10^{21.8}\rm{cm^{-2}}$.)
However and more importantly, short GRBs are consistent with the intrinsic absorption of long GRBs 
in the same redshift bin: A KS-test comparing the $NH_{\rm{HG}} $ distribution of long and short GRBs 
with $0 < z < 1$ reveals no statistical evidence for a distinct parent
population ($p(KS)=34\%$).

\subsection{$E_{\rm{pk}}-E_{\rm{\gamma,iso}}-E_{\rm{X,iso}}$ universal scaling}

\begin{figure}
\vskip -0.0 true cm
\centering
\includegraphics[scale=0.4]{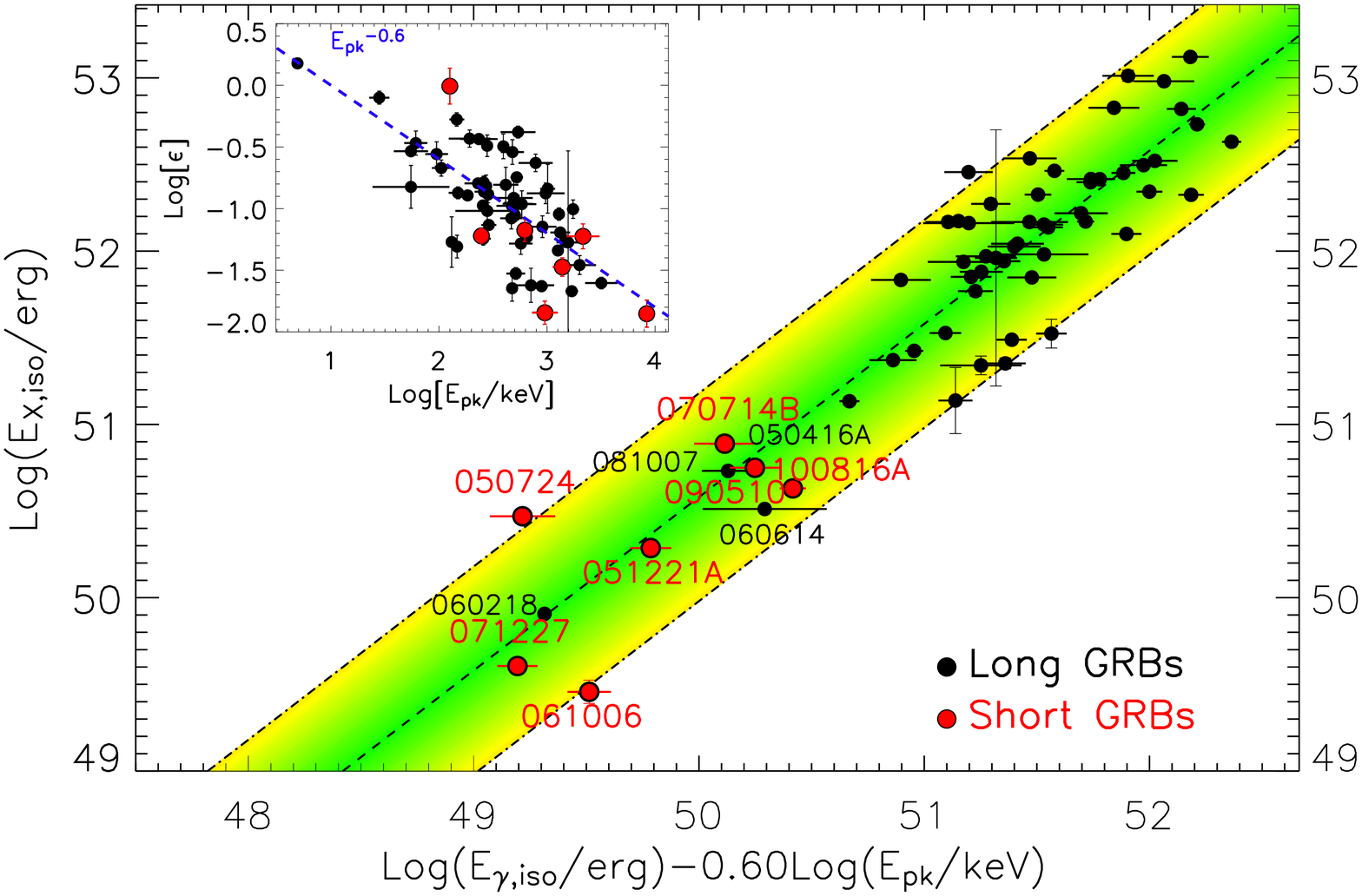}
\caption{Three parameter correlation involving $E_{\rm{pk}}$, $E_{\rm{\gamma,iso}}$ and 
$E_{\rm{X,iso}}$. Dashed line: best-fitting relation; dot-dashed lines
mark the 95\% confidence area around the best-fitting law. Notably, long and short GRBs share the same scaling, with short and sub-
energetic GRBs (like GRB 060218) occupying the same area of the plot. Inset: evolution of  
$\epsilon \equiv E_{\rm{X,iso}}/E_{\rm{\gamma,iso}}$ as
a function of the spectral peak energy of the prompt emission $E_{\rm{pk}}$.
Blue dashed line:  $\epsilon \propto E_{\rm{pk}}^{-0.6}$ scaling.}  
\label{Fig:3parcorr}
\end{figure}

Long and short GRBs are known to be \emph{not} consistent with the \emph{same} $E_{\rm{pk}}$ vs.
$E_{\rm{\gamma,iso}}$ scaling (where $E_{\rm{pk}}$ is the spectral peak energy during
the prompt emission, \cite{Amati08}); in this work we show that  this is also true
when we consider the $E_{\rm{X,iso}}$ vs. $E_{\rm{\gamma,iso}}$ correlation. However, 
when the three quantities are considered, we find evidence for a universal 
$E_{\rm{pk}}-E_{\rm{\gamma,iso}}-E_{\rm{X,iso}}$
scaling shared by ``normal" long GRBs, XRFs and short GRBs (Fig. \ref{Fig:3parcorr},
see also Bernardini et al., this volume). Computing the X-ray energy in the 0.3-30 (rest-frame)
energy band, the best fitting relation reads: 
\begin{equation}
	E_{\rm{X,iso}}\propto \frac{E_{\rm{\gamma,iso}}^{1.00\pm0.06}}{E_{\rm{pk}}^{0.60 \pm 0.10}}
\end{equation}
We note that: \\
(1) GRBs seem to divide into two groups with ``normal'' long GRBs occupying the upper-right area; 
short and peculiar GRBs together with XRFs share the same lower-left region of the plot.\\
(2) The scaling above implies: $E_{\rm{\gamma,iso}}/E_{\rm{X,iso}}\propto E_{\rm{pk}}^{0.60}$. The prompt
GRB efficiency reads: $\eta_{\gamma}=E_{\gamma}/(E_{\gamma}+E_{K})$, where $E_{K}$ is the
kinetic energy of the outflow. For $E_{\gamma}<E_{K}$, $\eta_{\gamma}\approx E_{\gamma}/E_{K}$.
Since $E_{K}$ is proportional to the X-ray energy in the afterglow,  we find $\eta_{\gamma} \propto
E_{\rm{pk}}^{0.60}$  (see the inset of Fig. \ref{Fig:3parcorr}, where we plot $\epsilon=1/\eta_{\gamma}$): 
GRBs with high spectral peak energy during the prompt emission tend to be more efficient.
This result has been shown to naturally arise in the context of the ``photospheric" emission model 
(Lazzati contribution, this volume; see also \cite{Fan12}). Alternatively, this correlation has also
been interpreted in the context of the ``cannon-ball" model: we refer the reader to \cite{Dado12} for details.

\section{Conclusions}
Our comprehensive statistical analysis of more than 650 GRBs (165 with measured redshift) allowed us to 
perform the first characterization of the X-ray afterglow of short GRBs as a class: short GRB afterglows
are less luminous, less energetic and show a faster decay. This study furthermore
revealed the existence of a universal GRB scaling. Interpreted in terms of prompt emission 
efficiency, this scaling may be used to shed light on the still elusive emission mechanism which powers the
prompt phase of GRBs as a whole.

\end{document}